\documentclass[conference]{IEEEtran}
\IEEEoverridecommandlockouts
\usepackage{cite}
\usepackage{amsmath,amssymb,amsfonts}
\usepackage{algorithmic}
\usepackage{graphicx}
\usepackage{textcomp}
\usepackage{xcolor}
\usepackage{multirow}
\usepackage{booktabs}
\usepackage{graphicx}
\usepackage{subcaption}
\def\BibTeX{{\rm B\kern-.05em{\sc i\kern-.025em b}\kern-.08em
    T\kern-.1667em\lower.7ex\hbox{E}\kern-.125emX}}
\begin{document}

\title{MoLEx: Mixture of LoRA Experts in Speech Self-Supervised Models for Audio Deepfake Detection\\
}

\author{
    \IEEEauthorblockN{Zihan Pan\IEEEauthorrefmark{1}, Sailor Hardik Bhupendra\IEEEauthorrefmark{1}, Jinyang Wu\IEEEauthorrefmark{1}}
    \IEEEauthorblockA{\IEEEauthorrefmark{1}Institute for Infocomm Research (I2R), Agency for Science, Technology and Research (A*STAR), Singapore, 138634\\
    Email: \{panz, Wu\_Jinyang\}@i2r.a-star.edu.sg, sailor.hardik2000@gmail.com}
    \thanks{Code available at: https://github.com/pandarialTJU/MOLEx-ORLoss}
    \thanks{This research is supported by the National Research Foundation, Prime Minister’s Office, Singapore, and the Ministry of Digital Development and Information, under its Online Trust and Safety (OTS) Research Programme (MDDI-OTS-001). Any opinions, findings and conclusions or recommendations expressed in this material are those of the author(s) and do not reflect the views of National Research Foundation, Prime Minister’s Office,  Singapore, or the Ministry of Digital Development and Information.}
}

\maketitle

\begin{abstract}

While self-supervised learning (SSL)-based models have boosted audio deepfake detection accuracy, fully finetuning them is computationally expensive. To address this, we propose a parameter-efficient framework that combines Low-Rank Adaptation with a Mixture-of-Experts router, called Mixture of LoRA Experts (MoLEx). It preserves pre-trained knowledge of SSL models while efficiently finetuning only selected experts, reducing training costs while maintaining robust performance. The observed utility of experts during inference shows the router reactivates the same experts for similar attacks but switches to other experts for novel spoofs, confirming MoLEx’s domain-aware adaptability. MoLEx additionally offers flexibility for domain adaptation by allowing extra experts to be trained without modifying the entire model. We mainly evaluate our approach on the ASVSpoof 5 dataset and achieve the state-of-the-art (SOTA) equal error rate (EER) of 5.56\% on the evaluation set without augmentation.
\end{abstract}


\section{Introduction}



The rapid advancement of speech synthesis, voice cloning, and voice conversion has greatly increased the risk of audio spoofing attacks, making it harder to distinguish deepfake audio from genuine speech \cite{wang24_ASVSpoof}. This poses serious threats to voice-based authentication systems, especially automatic speaker verification (ASV), which is now highly vulnerable to sophisticated spoofing techniques. The emergence of advanced synthesis engines, such as VALL-E, CosyVoice 2 \cite{du2024cosyvoice}, and GPT-4o, along with the accessibility of open-source and commercial deepfake tools, has further lowered the barrier for attackers. As a result, even individuals with minimal expertise can generate highly convincing synthetic voices, making reliable deepfake detection increasingly challenging.

Recent advances in deep learning have significantly improved audio deepfake detection, with models like AASIST3 \cite{borodin24_ASVSpoof} leveraging graph attention networks to enhance robustness. SSL models, such as Wav2Vec 2.0 \cite{baevski2020wav2vec}, HuBERT \cite{hsu2021hubert}, and WavLM \cite{chen2022wavlm} have also proven effective by capturing rich acoustic representations for spoof detection. Finetuning WavLM with specialized ResNet18 and self-attention backends has led to state-of-the-art results \cite{zhu24_ASVSpoof}. However, the high computational cost of fine-tuning these large-scale SSL models remains a challenge, highlighting the need for efficient architectures that balance performance with resource constraints.

This limitation underscores the need for parameter-efficient fine-tuning (PEFT) methods \cite{feng2023peft,lin2024peft}, which enable adaptation of large models without incurring the full cost of training all parameters. One widely adopted PEFT technique is LoRA \cite{zhang2023adalora,buehler2024x}, which reduces the memory footprint and training complexity by freezing the original SSL model weights and injecting trainable low-rank matrices into specific layers, such as dense layers \cite{hu2021lora}. In addition, scaling deep models efficiently for various spoofing attacks remains a challenge. To address this, Mixture of Experts (MoE) \cite{shazeer2017outrageously,he2024mixture,hyeon2024improving} introduces modular expert adapters, where only a subset of parameters is activated per input, significantly reducing the computational overhead of inference while maintaining strong representation learning. In this work, we aim to achieve both parameter-efficient and high-performance training for an SSL-based deepfake detection system by integrating LoRA with the MoE framework \cite{wang2022adamix,mehrish23_interspeech,kong2024moe,wumixture}. To this end, we propose Mixture of LoRA Experts (MoLEx), which introduces the following key contributions: 1) redesign the transformer \cite{vaswani2017attention} of SSL speech model and preserve the pre-trained knowledge. 2) propose a new loss function that enforces effective rank utilization for LoRA experts. 3) SOTA results in various ASVSpoof datasets, which contain the latest deepfake attacks. Besides, we also analyze the activation of the experts when confront different speaker pools or attacks. 4) extend the LoRA experts to new domains by only training new experts. 

The remainder of this paper is structured as follows: Section 2 introduces the MoLEx system and the orthogonality regularization loss function for LoRA. Section 3 outlines the experimental setup and discusses the evaluation of our proposed methods. In Section 4, we summarize and analyze the experimental results, highlighting the key findings and insights. Finally, Section 5 presents the conclusions.



\section{Methods}

\subsection{WavLM encoder, LoRA and MoE}

WavLM has shown strong performance across various speech processing tasks, including deepfake detection \cite{pan24c_interspeech}, by leveraging masked speech denoising and prediction to capture both content and speaker-related features\cite{chen2022wavlm}. However, finetuning such large-scale models remains computationally expensive \cite{rae2021scaling}. LoRA addresses this challenge by introducing low-rank trainable matrices into key layers while keeping the original model weights frozen, thereby cutting both memory and optimization footprints by an order of magnitude \cite{sundaram2019survey}. Although LoRA enables cheap adaptation, it alone cannot exploit input-dependent specialization. MoE addresses that gap by routing each sample to just a few expert sub-networks, achieving near-linear scaling of model capacity with constant inference cost \cite{zhou2022mixture}. In deepfake detection, MoE enhances adaptability by assigning different experts to detect specific spoofing techniques. Integrating LoRA and MoE offers a powerful parameter—efficient finetuning with selective expert activation, making large-scale deepfake detection both scalable and computationally feasible.
\begin{figure*}[t]
  \centering
  \includegraphics[width=0.75\linewidth]{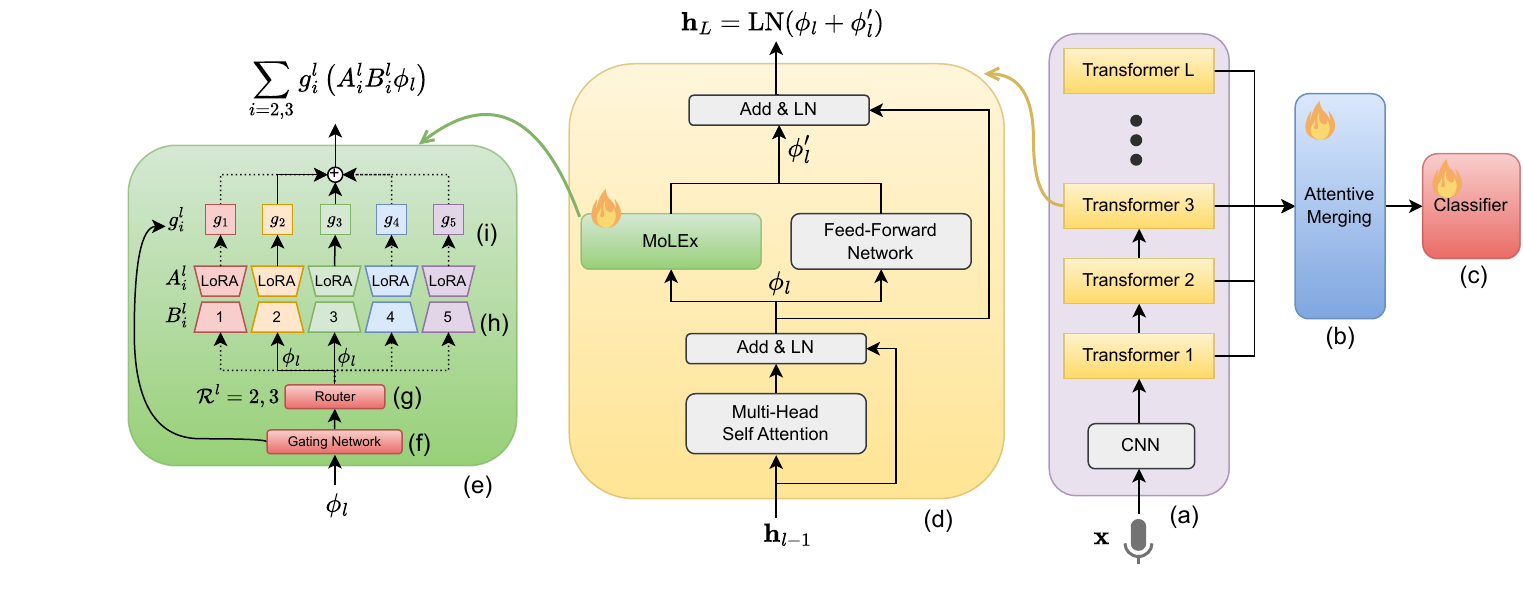}
  \caption{Diagram of the MoLEx in speech SSL model for audio deepfake detection. Grey colored blocks are frozen during training.}
  \label{fig:method_block}
\end{figure*}
\subsection{Proposed Mixture of LoRA Experts (MoLEx) for the transformer layers in WavLM}
We propose the idea of MoLEx as shown in Fig. \ref{fig:method_block}: 1) We modify the transformer layers with MoLEx in a pre-trained WavLM model in Fig. \ref{fig:method_block}(a)(d), and use it in an audio deepfake detection system. 2) $N$ LoRA adapters in Fig. \ref{fig:method_block}(e)(h) are placed in parallel with the dense feed-forward network in a transformer layer in Fig. \ref{fig:method_block}(d). 3) the $N$ LoRA adapters serve as experts, and a gating network in Fig. \ref{fig:method_block}(f) is trained to select a subset of $K$ experts for activation during training and inference.
Then we formulate the framework as follows: the speech encoder of the SSL pre-trained WavLM model in Fig. \ref{fig:method_block}(a) consists of Convolutional Neural Network (CNN) layers and stacks of transformer layers.
The CNN extracts an initial acoustic representation from the input speech sequence $\mathbf{x}$:
\begin{equation} \mathbf{h}_0 = \text{CNN}(\mathbf{x}) \end{equation}
The extracted feature $\mathbf{h}_0$ is processed through multiple transformer layers. Each transformer layer \( l \) consists of two main components: Multi-Head Self-Attention (MHSA) and Feed-Forward Network (FFN). For the $l^{th}$ transformer layer, we denote the output of the MHSA (with layer normalization and residual connection) as $\phi_l$, which is further fed into both the MoLEx and FFN layers in parallel. 
Instead of updating the full weight matrix \( W^l \) in each FFN layer, the LoRA introduces a low-rank decomposition:
\begin{equation}
W^l = W^l_{\text{FNN}} + W^l_{\text{LoRA}}, \quad \text{where} \quad W^l_{\text{LoRA}} = A^l B^l
\end{equation}
\( W^l_{\text{FNN}}  \) is the pre-trained weight matrix in FNN, which remains frozen during training. \( W^l_{\text{LoRA}} \) is the trainable adaptation component, using two low-rank matrices \( A^l \in \mathbb{R}^{d \times r} \) and \( B^l \in \mathbb{R}^{r \times d} \) with \( r \ll d \), reducing the number of trainable parameters.
Instead of a single LoRA-adapted weight matrix, we use a Mixture of LoRA Experts (MoLEx) framework, where $N$ multiple LoRAs as experts are trained, and a noisy gating mechanism \cite{fedus2022switch} selects the best combination of $K$ experts dynamically.
A noisy gating network \cite{shazeer2017outrageously} (Fig. \ref{fig:method_block}(f)) determines the contribution of each expert:
\begin{equation}
g_i^{l} = \text{softmax}(W_G^{l} \mathbf{\phi}_l), \quad i=1,2,...,N
\label{eq: gating scores}
\end{equation}
where $g_i^{l}$ is the gating score for expert $i$ in the $l^{th}$ layer. $W_G^{l}$ is the linear gating network that computes attention scores based on the input representation $\mathbf{\phi}_l$. To ensure computational efficiency, the router (Fig. \ref{fig:method_block}(g)) selects only the top $K$ experts based on the highest gating scores, which is defined as:
\begin{equation} 
    \mathcal{R}^l = \arg \operatorname{topK} (\mathbf{g}^{l}, K) = { i_1, i_2, \dots, i_K, \quad K\leq N } 
\end{equation}
where $g_{i_1}^{l} \geq g_{i_2}^{l} \geq \dots \geq g_{i_K}^{l}$. The router $\mathcal{R}^l$ returns the set of indices $i_1, i_2, \dots, i_K$ corresponding to the top $K$ highest values from $g^l_i$.
The final output $\mathbf{\phi}_l'$ from the MoLEx and FNN is computed from a weighted sum of top $K$ LoRA experts:
\begin{equation}
\mathbf{\phi}_l' = \sum_{i=i_1}^{i_K} g_i^{l} \left(A_i^{l} B_i^{l} \mathbf{\phi}_l\right) + W^l_{\text{FNN}} \phi^{l}
\end{equation}
$g_i^{l}$ dynamically controls the influence of each LoRA expert. This MoLEx approach allows the model to adaptively switch between different specialized LoRA experts to improve generalization. The final output of the $l^{th}$ transformer layer is then defined as:
\begin{equation}
    h^l = \text{LN}(\phi_l + \mathbf{\phi}_l')
\end{equation}
where LN denotes layer normalization to stabilize training. 

To aggregate embeddings $\phi^l$ from multiple transformer layers, an attentive merging mechanism in Fig. \ref{fig:method_block}(b) is applied, which is proposed and evaluated in \cite{pan24c_interspeech,guragain2024speech}. Finally, the $L$ merged embeddings are fed into a classifier in Fig. \ref{fig:method_block}(c) for distinguishing spoof speech from bonafide.

For concreteness we report parameter counts of the configuration used for the best performance on ASVSpoof 5 dataset: a WavLM-Large backbone (using 12 transformer layers) with 12 LoRA experts (rank $r=32$) per layer. Full end-to-end fine-tuning of this stack would introduce $376,442,578$ trainable parameters. Instead, MoLEx freezes the backbone and updates only the three light-weight components highlighted in Fig.\ref{fig:method_block}, namely the MoLEx (e), the attentive-merging module (b), and the classifier (c), for a combined $51,275,794$ parameters. This corresponds to an 86.4\% reduction of parameters.


\subsection{Orthogonality regularization loss for MoLEx}
\label{sec: Ortho loss}
With the rank $r$ of $A$ and $B$ constrained, $W_{\text{LoRA}}$ may lose representational capacity since $\text{rank}(A B) \leq r$. To enhance the expressiveness of LoRA experts while maintaining a low-rank constraint \cite{zeng2023expressive}, we introduce the orthogonality regularization loss function $\mathcal{L}_{\text{orth}}$ that enforces the up and down projection matrices $A$ and $B$ to approximate a maximally expressive low-rank transformation. We first define the Gram matrix of $AB$ as:
\begin{equation}
G = (A B)(A B)^T
\end{equation}
To enforce the orthogonality of each vector in the Gram matrix, we define the $\mathcal{L}_{\text{orth}}$ as the summation of the Mean Squared Error (MSE) loss between $G$ and the identity matrix $I$ across all experts:
\begin{equation}
\mathcal{L}_{\text{orth}} = \sum_{m=1}^{\mathcal{M}} \sum_{i=i_1}^{i_K} ||G^m_i-I||_{F^2}
\end{equation}
where $||\cdot ||_F$ and $\mathcal{M}$ denote the Frobenius norm and number of MoLEx modules, respectively. By updating the model with $\mathcal{L}_{\text{orth}}$, we ensure that $W^l_{\text{LoRA}}$ remains as close as possible to an fully allocated rank transformation, thus maintaining expressiveness while adhering to a low-rank constraint.
\begin{figure}[t]
  \centering
  \includegraphics[width=0.65\linewidth]{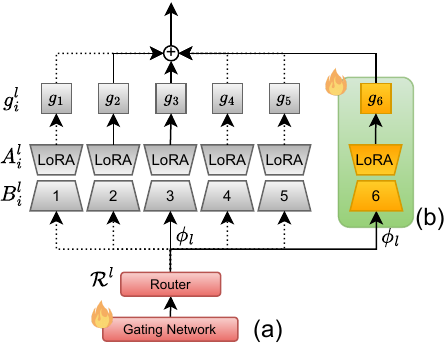}
  \caption{Illustration of extending MoLEx module. Only new expert 6 and the router are updated, while the other grey colored experts are frozen during training on new dataset.}
  \label{fig:lora extension}
\end{figure}
\subsection{Extension of MoLEx for domain adaptation}
Due to the flexibility of MoLEx, we can easily adapt a trained model to new dataset domains by adding new LoRA experts.  Fig. \ref{fig:lora extension} illustrates how to extend a MoLEx with new experts. For instance, the $6^{th}$ LoRA expert in Fig. \ref{fig:lora extension}(b) is added in parallel to the five existing experts. During adaptation to new domains, only the newly added $6^{th}$ expert and the router (Fig. \ref{fig:lora extension}(a)) in MoLEx will be updated, while all other parameters in the corresponding transformer layer will be frozen.
\section{Experiment setup}
\subsection{Training and evaluation dataset}
We mainly conduct the ablation study and evaluate our proposal on the ASVSpoof 5 challenge dataset \cite{ASVSpoof5}, which is designed to promote research on speech spoofing, deepfake detection, and adversarial attacks. Compared to previous challenges, ASVSpoof 5 introduces a more diverse dataset, featuring crowd-sourced data from thousands of speakers recorded in real-world conditions, as well as stronger spoofing and adversarial attacks optimized to deceive countermeasure (CM) systems. The dataset includes speech deepfake attacks generated with state-of-the-art text-to-speech (TTS) and voice conversion (VC) models, such as GlowTTS, GradTTS, VITS, StarGANv2-VC, and YourTTS \cite{wang24_ASVSpoof}. In this work, we evaluate our deepfake detection system on counter-measure (CM) evaluation Track 1 only. Table \ref{tab:ASVSpoof5_stats} summarizes the number of spoofed and bonafied samples in various sets. In addition, we also evaluate the proposal on other main stream deepfake detection datasets: ASVSpoof 2019, ASVSpoof 2021LA/DF, DFADD, FakeOrREal, and In the Wild.  Besides, we also conduct domain adaptation experiments on ASVSpoof 2019LA, 2021LA, and 2021DF.
\begin{table}[h]
    \centering
    \caption{ASVSpoof 5 Dataset statistics.}
    \setlength{\tabcolsep}{0.1mm}
    \renewcommand{\arraystretch}{1.0} 
    \begin{tabular}{lccc}
        \toprule
        \textbf{Partition set} & \textbf{Bonafide} &\quad \textbf{Spoofed} &\quad \textbf{Total} \\
        \midrule
        Training & 18,797 & 163,560 & 182,357 \\
        Development & 31,334 & 109,616 & 140,950 \\
        Evaluation (Track 1) & 138,688 & 542,086 & 680,774 \\
        \bottomrule
    \end{tabular}
    \label{tab:ASVSpoof5_stats}
\end{table}

\subsection{Models and tasks}
The diagram of our proposed system is demonstrated in Fig. \ref{fig:method_block}(a)(b)(c). We use the SSL pre-trained WavLM large model as the speech foundation model. The attentive merging model follows the settings in \cite{pan24c_interspeech,guragain2024speech}, in which they found that merging the hidden embeddings from the early transformer layers in the WavLM model contributes significantly to anti-spoofing task. So we set the number of transformer layers $L=12$ in most of our experiments. The classifier is a combination of one long-short term memory (LSTM) layer with hidden dimension of $192$, and one fully-connected layer with output dimension of $2$. Among all model weights, only the MoLEx modules, attentive merging block and classifier will be updated during training. 

We first investigate the effectiveness of the orthogonality regularization loss $\mathcal{L}_\text{orth}$ by checking the effective rank of experts after training. We train the MoLEx framework by minimizing two settings of loss functions: 1) cross-entropy loss $\mathcal{L}_\text{ce}$; 2) $\mathcal{L}_\text{orth}+\mathcal{L}_\text{ce}$. In linear algebra \cite{stanfordSVD}, the rank of a real matrix is equal to the number of non-zero singular values by singular value decomposition (SVD). The effective rank $\text{rank}_{\tau}(AB)$ of a LoRA expert $AB$ is derived by the number of dominate singular values exceeding a tolerance threshold $\tau$ \cite{zeng2023expressive}: 
\begin{equation}
    \text{rank}_{\tau}(AB) = \sum_i 1 (\sigma_i(AB) \ge \tau)
\end{equation}
where $\sigma_i(AB)$ represents the singular values of matrix $AB$; $\tau$ filters out weaker singular values, ensuring that only the dominant components contribute. This promotes an effective orthogonal transformation, allowing each expert matrix to maintain maximal expressiveness within its rank constraint.

Next, we evaluate our proposed system with various hyperparameter settings on the latest ASVSpoof 5 dataset, followed by a comparison with other benchmark models. Then we fix the optimal hyperparameter setting and evaluate the model on other audio deepfake datasets. Besides the EER numbers, we will investigate how the trained experts are triggered during inferring various spoof attack input $\mathbf{x}$, by averaging the gating scores $g_i^l$ assigned to the $i^{\text{th}}$ expert from the $l^{\text{th}}$ MoLEx layer over each dataset:
\begin{equation}
    \mathcal{G}_i^l=\frac{1}{N_{\mathbf{X}}}\sum_{\mathbf{x}}{g_i^l(\mathbf{x})}
\label{eq: expert utilize score}
\end{equation}
$g_i^l$ is derived from Eq.\ref{eq: gating scores} and illutrated in Fig.\ref{fig:method_block}(e). $N_{\mathbf{X}}$ denotes the total number of input samples from the dataset. We define $\mathcal{G}_i^l$ as the expert utilization score, whose higher values indicate more contribution from the corresponding expert. 

Finally, we conduct a domain adaptation experiment to demonstrate the adaptability of MoLEx to new datasets. Starting with a model pre-trained on the ASVSpoof 5 training set, we introduce additional experts to the MoLEx modules and finetune the model on the ASVSpoof 2019LA training set. During this process, only the newly added experts and the routers are updated, while the original model remains unchanged. We evaluate the adapted model on multiple evaluation sets, including ASVSpoof 5, to assess its generalization capability.



\section{Results and discussion}

\begin{figure*}[t]
  \centering
  \begin{subfigure}[b]{0.21\textwidth}
    \includegraphics[width=\linewidth]{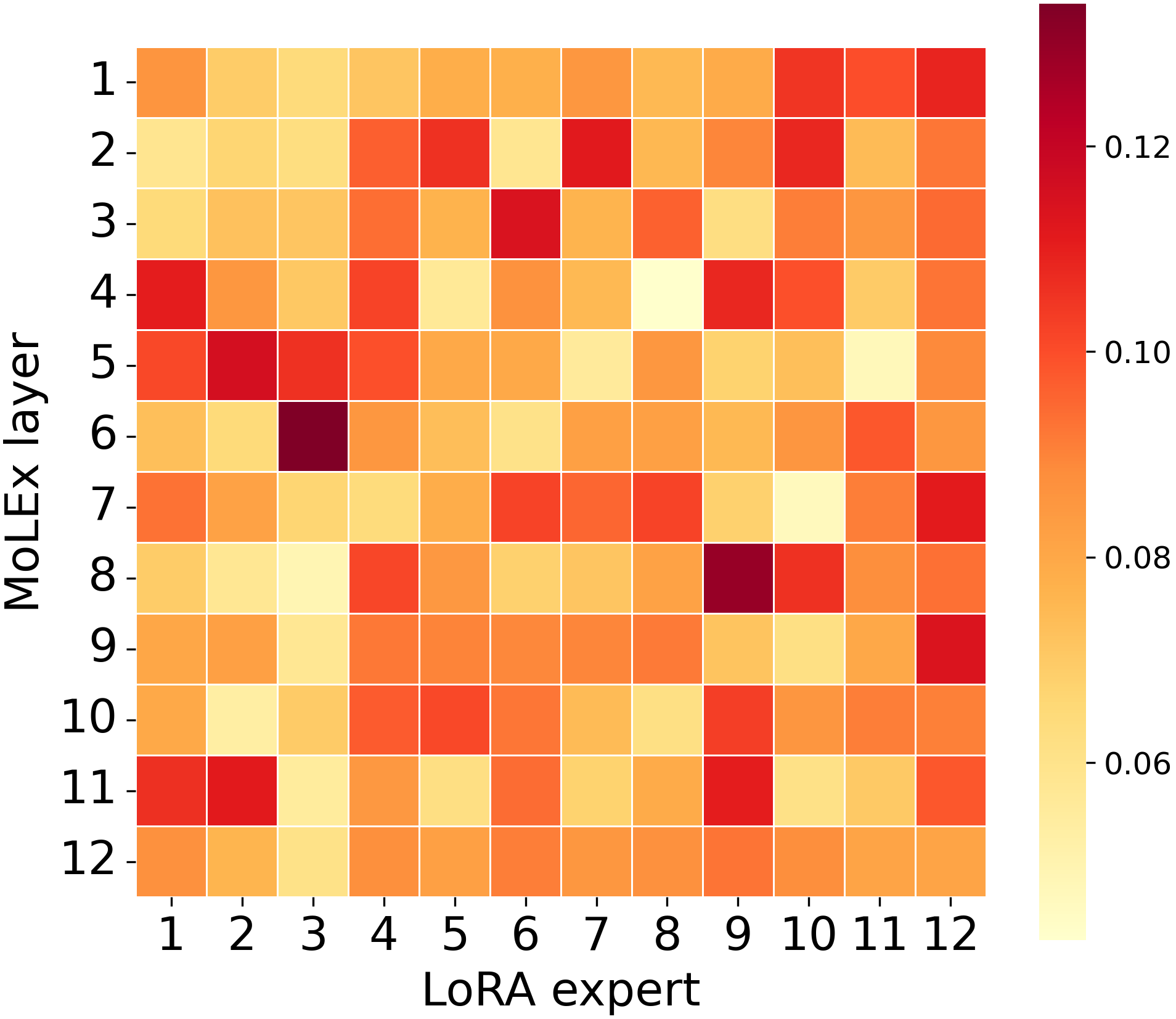}
    \caption{ASVSpoof 2019}
  \end{subfigure}
  \begin{subfigure}[b]{0.21\textwidth}
    \includegraphics[width=\linewidth]{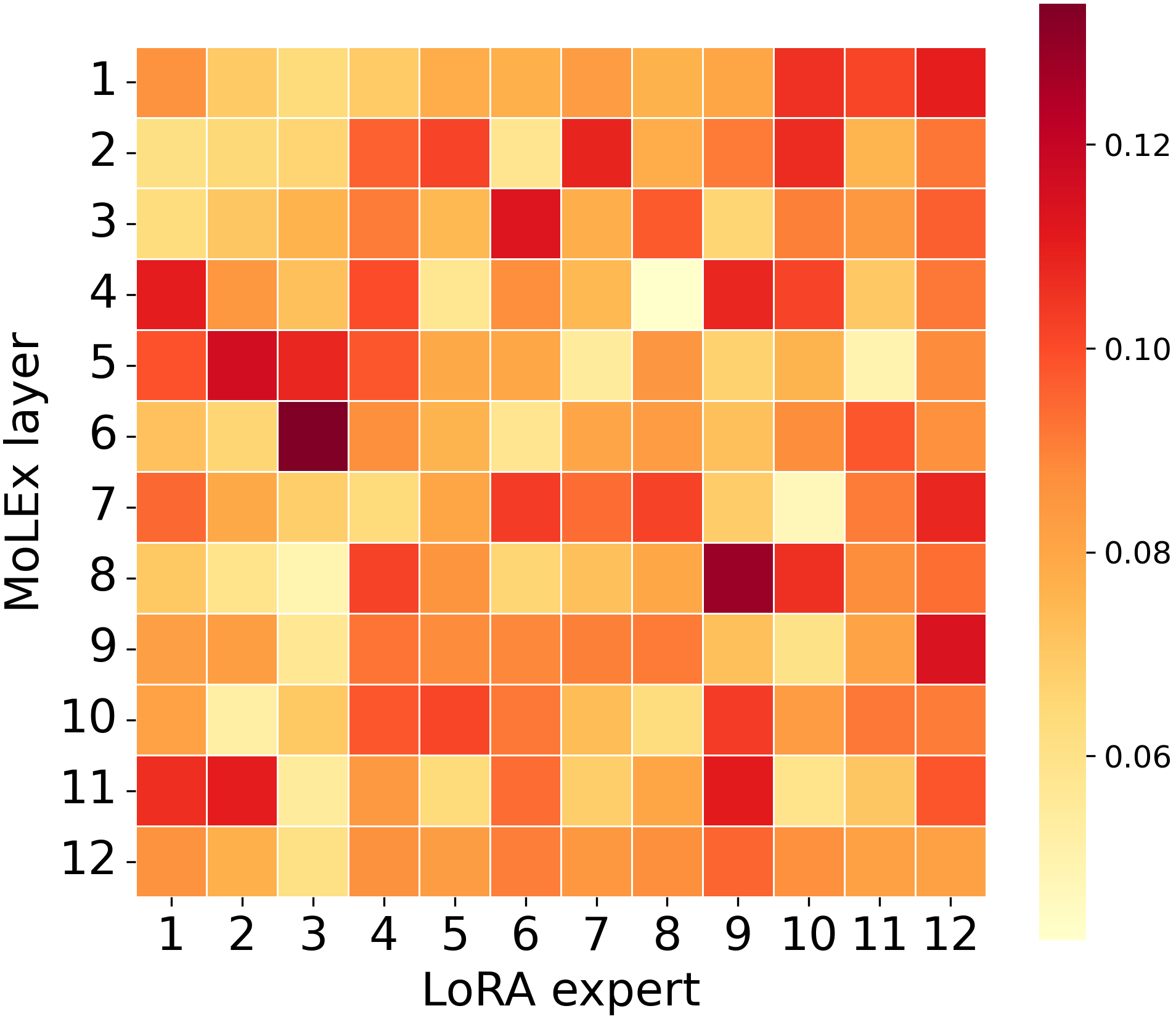}
    \caption{ASVSpoof 2021 LA}
  \end{subfigure}
  \begin{subfigure}[b]{0.21\textwidth}
    \includegraphics[width=\linewidth]{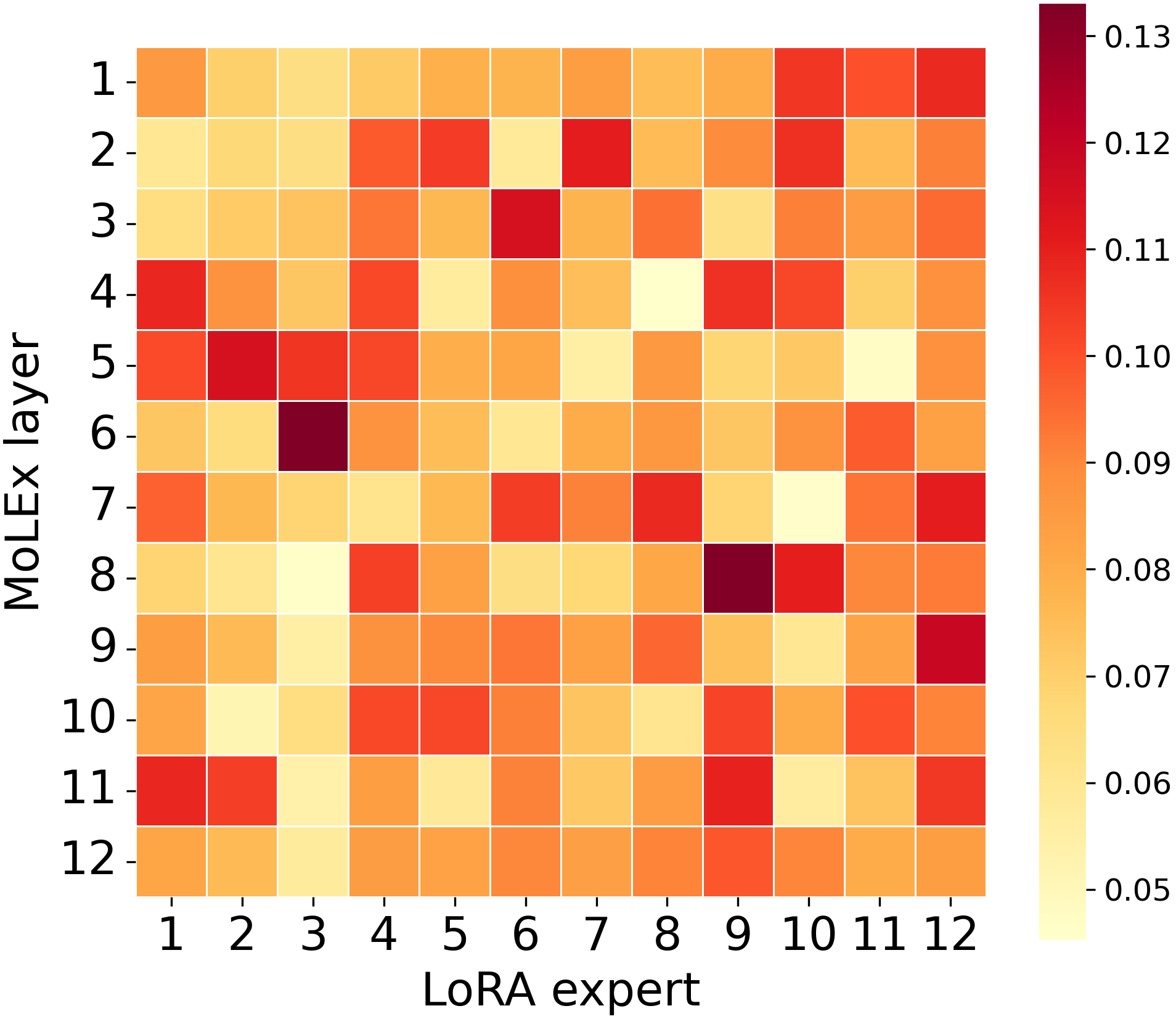}
    \caption{ASVSpoof 2021 DF}
  \end{subfigure}
  \begin{subfigure}[b]{0.21\textwidth}
    \includegraphics[width=\linewidth]{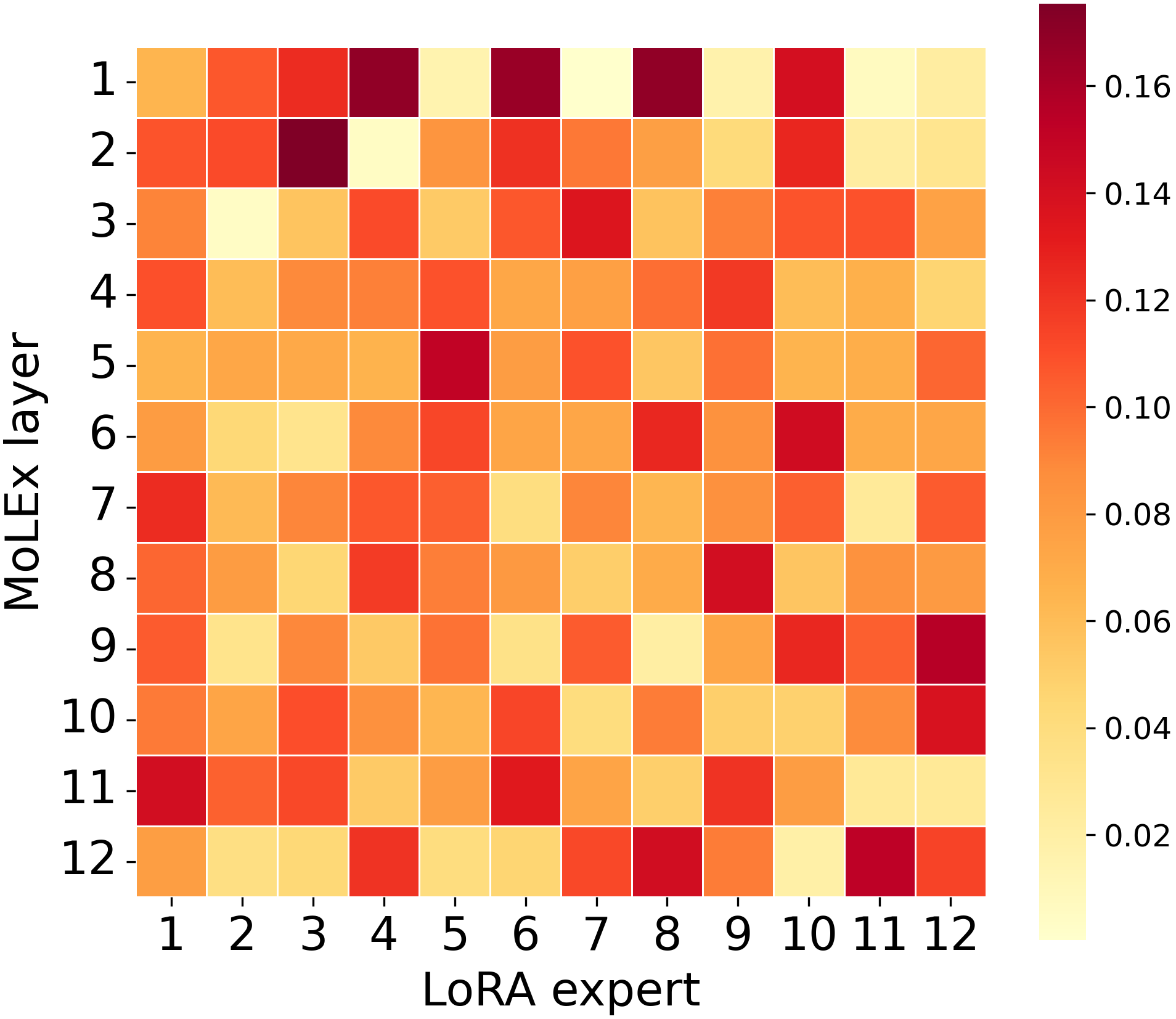}
    \caption{ASVSpoof 5}
  \end{subfigure}
  
  
  \begin{subfigure}[b]{0.21\textwidth}
    \includegraphics[width=\linewidth]{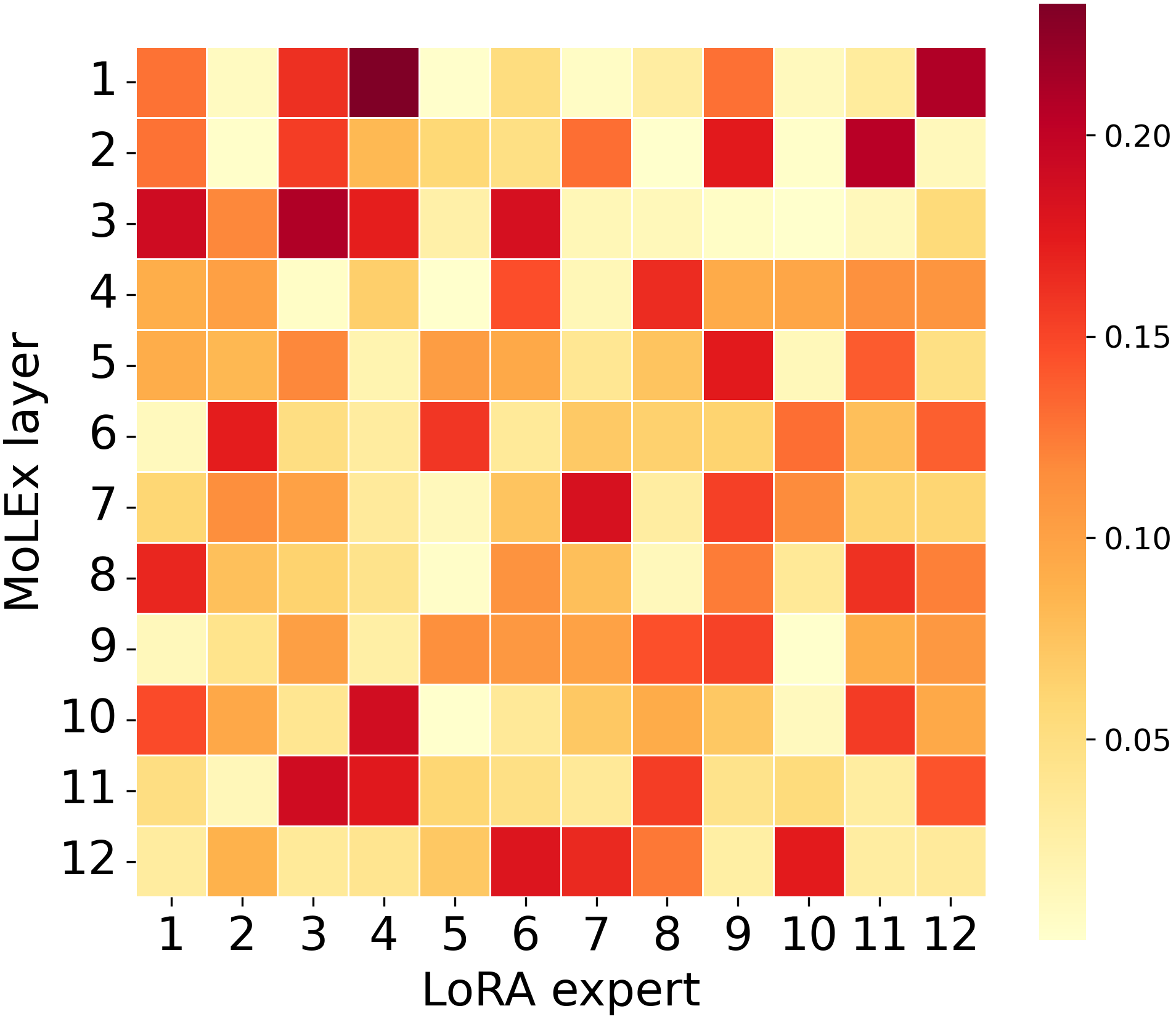}
    \caption{DFADD}
  \end{subfigure}
  \begin{subfigure}[b]{0.21\textwidth}
    \includegraphics[width=\linewidth]{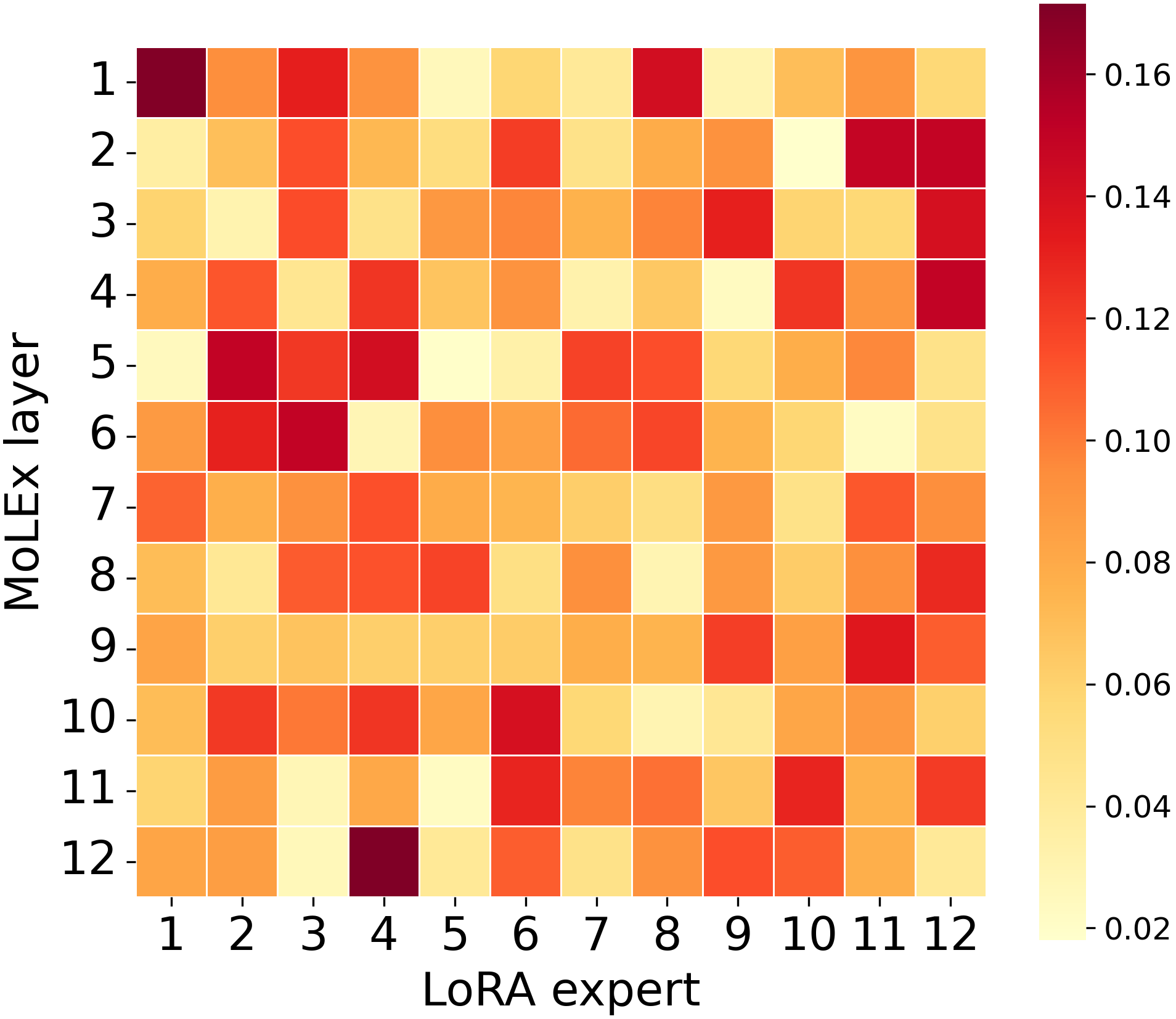}
    \caption{Fake Or Real}
  \end{subfigure}
  \begin{subfigure}[b]{0.21\textwidth}
    \includegraphics[width=\linewidth]{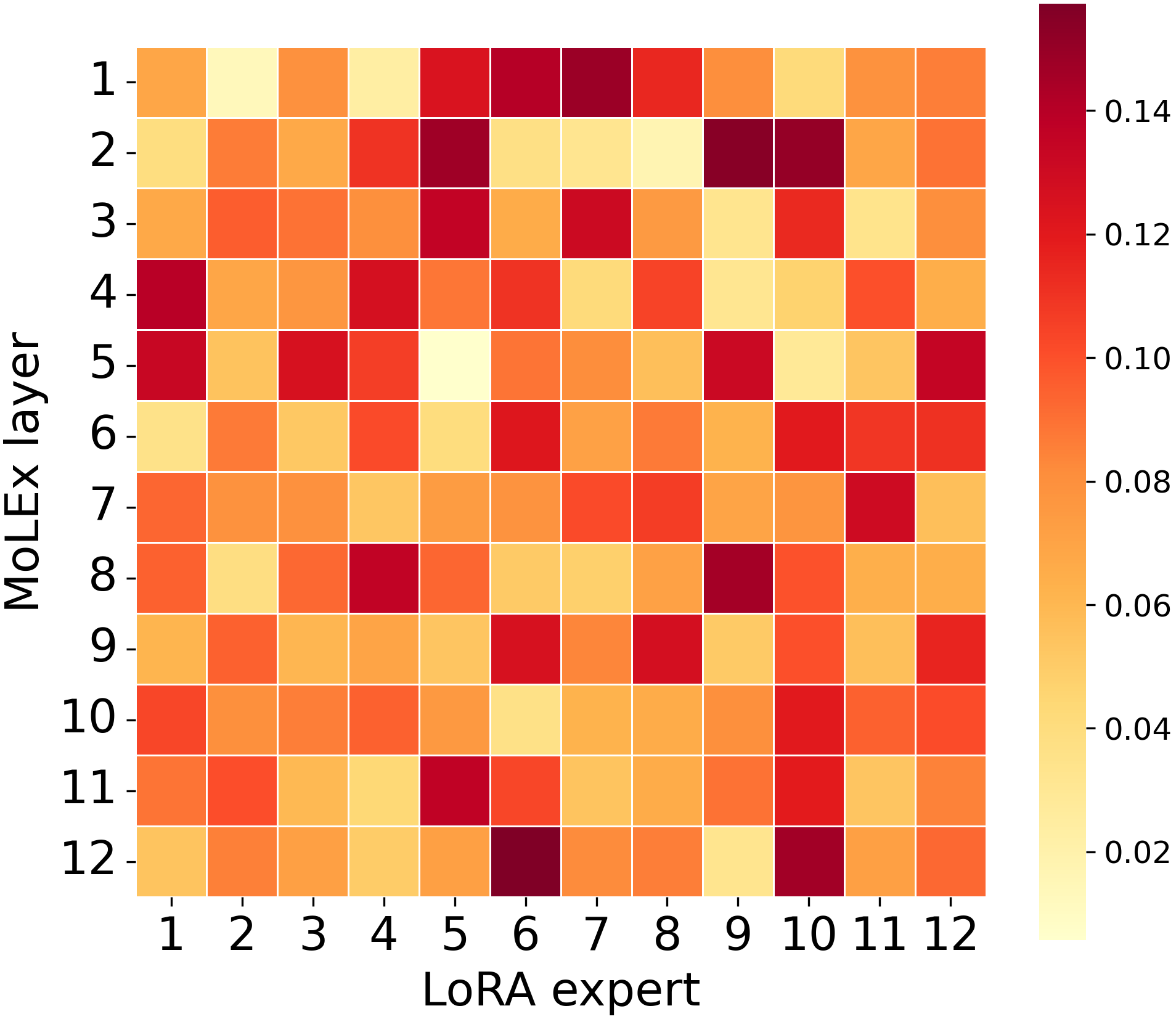}
    \caption{In The Wild}
  \end{subfigure}
  \begin{subfigure}[b]{0.21\textwidth}
    \includegraphics[width=\linewidth]{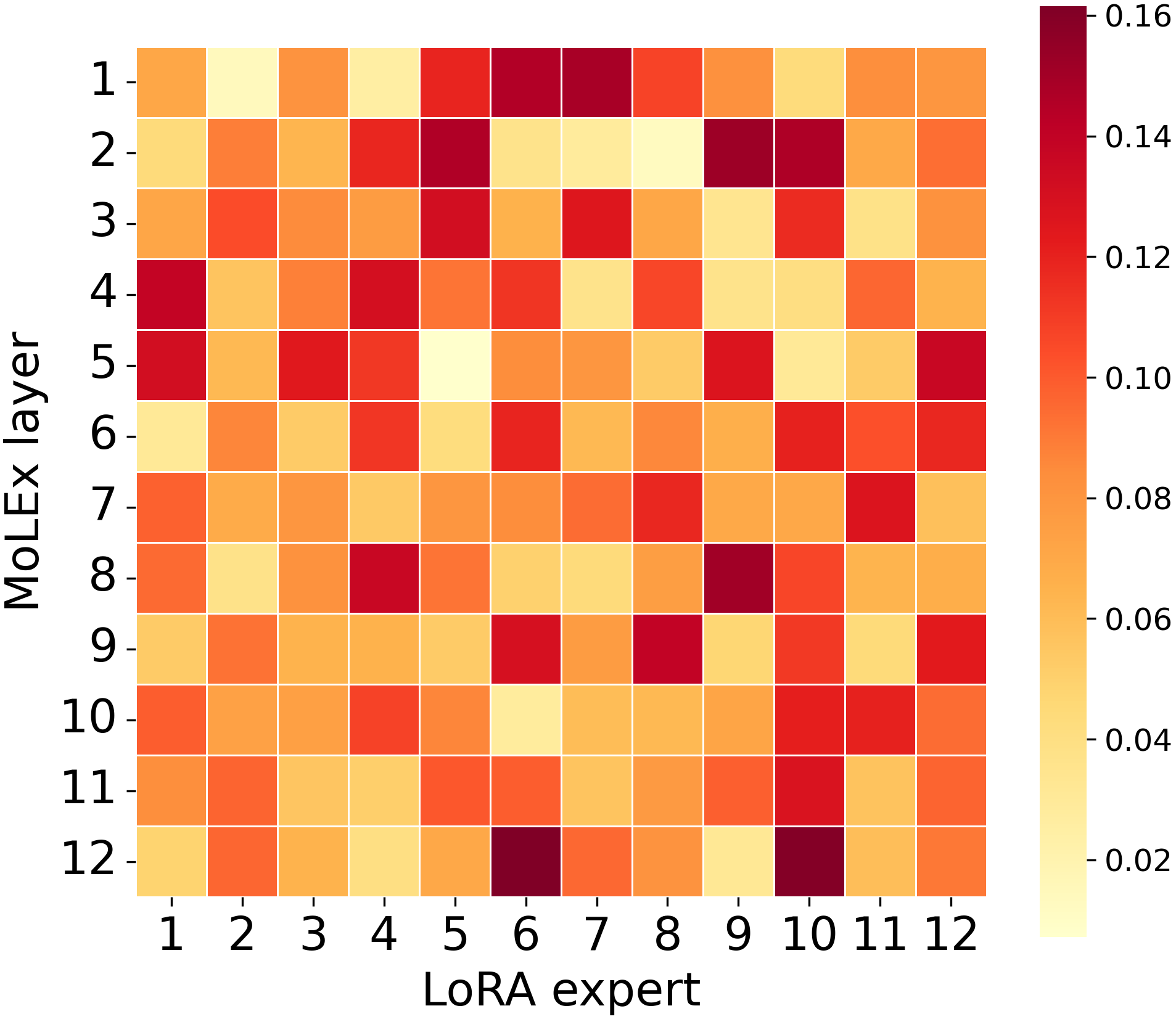}
    \caption{LibriVoc}
  \end{subfigure}

  \caption{The utilization scores for the cross-data evaluation experiment in Table \ref{tab:train_eval_matrix}. Each subfigure illustrates the average gating scores at inferring the corresponding evaluation set. The x-axis and y-axis of each subfigure represent the MoLEx layer and its experts, respectively. Each value in the subfigure is derived from $\mathcal{G}_i^l$ in Eq.\ref{eq: expert utilize score}}.
  \label{fig:all_8_heatmaps}
\end{figure*}

\subsection{LoRA experts rank}


Table \ref{tab: LoRA rank investigate} presents the average effective rank of all LoRA experts after training. The experts trained with \(\mathcal{L}_{\text{ortho}}\) maintain effective rank utilization, whereas those trained without it experience significant rank degradation, particularly when \(\tau=10^{-2}\). This suggests that when too many singular values become small, the experts are underutilized, limiting their expressiveness. Enforcing a minimum rank through \(\tau\) ensures that experts fully utilize their allocated capacity, preventing rank collapse. Consequently, maintaining a maximally expressive low-rank transformation with a larger \(\tau\) enhances their representation power and task performance.

\begin{table}[t]
\centering
\caption{The average effective rank of experts after training with various loss functions.}
\scriptsize
\footnotesize
\setlength{\tabcolsep}{0.1mm}
\renewcommand{\arraystretch}{1.0} 
\begin{tabular}{lcccccc}
\hline
\toprule
\multicolumn{1}{l}{\textbf{Loss type}} & \multicolumn{3}{l}{\quad \textbf{$\mathcal{L}_{\text{orth}}+ \mathcal{L}_{\text{ce}}$} \quad} & \multicolumn{3}{c}{\textbf{$\mathcal{L}_{\text{ce}}$}}\\
\hline
Allocated rank & \quad 64 & \quad 32 & \quad16  \qquad \qquad& \quad 64 &  \quad32  & \quad16 \\
\hline
$\tau=10^{-2}$ & \quad 64 & \quad 32 & \quad16 \qquad \qquad & \quad 18  &  \quad7  & \quad4\\
$\tau=10^{-3}$ & \quad 64 & \quad 32 & \quad16  \qquad \qquad & \quad 55  &  \quad26  & \quad13\\
\hline
\bottomrule
\end{tabular}
\label{tab: LoRA rank investigate}
\end{table}
\subsection{Evaluation results on ASVSpoof 5 dataset}
\begin{table}[t]
\centering
\caption{Evaluation performance on ASVSpoof 5 Track 1. (dev-devlopment set; eval-evaluation track 1 set)}
\footnotesize
\setlength{\tabcolsep}{0.6mm}{
\renewcommand{\arraystretch}{1.0}{
\begin{tabular}{lcccccc|rr}
\toprule
\multicolumn{7}{c}{\textbf{Model configuration}} & \multicolumn{2}{r}{\textbf{EER (\%)}} \\
\midrule
  & $L$& $\mathcal{M}$ & $K/N$ & $r$ \quad & $\mathcal{L}_{\text{ce}} $ & $\mathcal{L}_{\text{orth}} $  & \quad dev  &  \quad eval \\
\hline
\midrule
 & 12 & 12 & 4/12 & 32 \qquad & $\checkmark$ & $\times$ & 3.57 & 16.92 \\
 & 12 & 12 & 4/12 & 16\qquad & $\checkmark$ & $\times$ & 2.96 & 7.75 \\
 & 12 & 12 & 4/12 & 8\qquad & $\checkmark$ & $\times$ & 2.78 & 7.04 \\
 & 12 & 12 & 3/9 & 32\qquad & $\checkmark$ & $\checkmark$ & 4.18 & 8.16 \\
 & 12 & 12 & 2/6 & 32\qquad & $\checkmark$ & $\checkmark$ & 0.78 & 6.30 \\
  & 12 & 12 & 2/12 & 32 \qquad& $\checkmark$ & $\checkmark$ & 0.37 & 7.60 \\
  & 12 & 12 & 4/12 & 32 \qquad& $\checkmark$ & $\checkmark$ & 1.25 & 5.56 \\
 & 12 & 12 & 8/12 & 32\qquad & $\checkmark$ & $\checkmark$ & 0.78 & 5.69 \\
 & 12 & 12 & 12/12 & 32 \qquad& $\checkmark$ & $\checkmark$ & 0.55 & 5.78 \\
 & 12 & 12 & 4/12 & 16 \qquad& $\checkmark$ & $\checkmark$ & 0.49 & 6.80 \\
  & 12 & 12 & 4/12 & 8 \qquad& $\checkmark$ & $\checkmark$ & 0.85 & 7.51 \\

  & 12 & 6 & 4/12 & 32 \qquad& $\checkmark$ & $\checkmark$ & 1.95 & 7.23 \\
    
      & 12 & - & - & - \qquad& $\checkmark$ & $\times$ & 1.98 & 9.62 \\
      & 24 & - & - & - \qquad& $\checkmark$ & $\times$ & 10.83 & 15.82 \\
        & 12 & - & 1/- & 32 \qquad& $\checkmark$ & $\times$ & 6.43 & 15.44 \\   
                    
\hline
\bottomrule
\hline
\end{tabular}
}}
\label{tab:ASVSpoof5}
\end{table}
The experiments are typically conducted using a forzen WavLM large model with transformer layer index $l=1,2,...,L, \quad L=12$, with varying the expert rank \( r \), the number of MoLEx modules $\mathcal{M}$, and the top-\( K \) expert selection out of total $N$ experts. We also assess the impact of including the orthogonality regularization loss \(\mathcal{L}_{\text{orth}}\) alongside the cross-entropy loss \(\mathcal{L}_{\text{ce}}\). Baseline comparisons include: 1) standard WavLM large model without MoLEx, with 12-layer and 24-layer variants; 2) WavLM large model with sigle expert for each transformer, assessing performance without mixture-based selection.
\begin{table}[t]
\centering
\caption{Performance from other single models on ASVSpoof 5 evaluation Track 1. (finetuned-FT; data augmented-DA)}
\footnotesize
\setlength{\tabcolsep}{0.6mm}{
\renewcommand{\arraystretch}{1.0}{
\begin{tabular}{llrr}
\hline
\toprule
\multicolumn{2}{c}{\textbf{Model name}} & \multicolumn{2}{c}{\textbf{EER (\%)}} \\
\cline{3-4}
\multicolumn{2}{c}{} & \textbf{dev} & \textbf{eval}\\
\hline
\midrule
\multirow{10}{*}{\rotatebox{90}{Public}} & Wav2Vec2-AASIST-KAN \cite{borodin24_ASVSpoof} & - & 22.67  \\
                         & SEMAA-1 \cite{xia24_ASVSpoof} & 14.93 & 23.63 \\
                         & AASIST-CAM++ fused \cite{truong24_ASVSpoof}& 10.48 & 25.47 \\                         
                         & WavLM-AASIST \cite{chan24_ASVSpoof}& 12.18 & - \\
                         & WavLM-LCNN9 \cite{chan24_ASVSpoof}& 15.81 &-  \\
                         & WavLM-ResNet18-SA \cite{chan24_ASVSpoof}& 7.04 & - \\
                         & WavLM-MHFA \cite{stourbe24_ASVSpoof}& 6.78 & -\\
                         & WavLM (FT-DA) \cite{combei24_ASVSpoof} & 6.56 & 17.08\\
                         & Best submission in challenge \cite{ASVSpoof5} & - & 8.61 \\
                         & WavLM-RawBoost (FT-DA) \cite{zhu24_ASVSpoof} & 3.0 & 5.5\\
                    
\hline
 & Our WavLM-MoLEx-LSTM & 1.25 & 5.56\\
\hline
\bottomrule
\hline
\end{tabular}
}}
\label{tab:compare_results}
\end{table}

Table \ref{tab:ASVSpoof5} summarizes the results on the ASVSpoof 5 evaluation track 1 set. Key findings include:  
1) effect of expert rank (\( r \)): A lower LoRA rank significantly impacts performance. When \( r = 8 \), the EER degrades from 5.56\% to 7.51\%, demonstrating the importance of maintaining an adequate rank space for adaptation. 2) Impact of top-\( K \) selection: selecting fewer experts per input (e.g., \( K/N = 2/6 \)) improves efficiency while maintaining high detection accuracy. The best performance (5.56\% EER) is achieved with \( K/N = 4/12 \). However, selecting too many experts (e.g., \( K/N = 12/12 \)) slightly worsens performance due to redundant activation. 3) Orthogonality regularization loss (\(\mathcal{L}_{\text{orth}}\)) matters: adding \(\mathcal{L}_{\text{orth}}\) improves performance across settings. Without it, the best EER is 7.04\%, whereas 5.56\% is achieved with the full MoLEx setting. This confirms that enforcing rank constraints on LoRA experts enhances expressiveness. 4) Comparison with baselines: standard non-MoLEx models perform significantly worse, with 9.62\% EER (12-layer SSL) and 15.82\% EER (24-layer SSL). This highlights the effectiveness of MoLEx's expert selection strategy over traditional finetuning. Our best result also outperforms other single models reported from recent papers in Table \ref{tab:compare_results}. Specifically, we achieve comparable performance to a finetuned WavLM model with data augmentation \cite{zhu24_ASVSpoof}. These results demonstrate that MoLEx provides a balance between parameter efficiency and performance effectiveness, significantly improving deepfake detection capability.

\subsection{Evaluate on additional benchmarks}

\begin{table}[t]
  \renewcommand{\arraystretch}{0.85} 
  \centering
  \caption{Cross-dataset evaluation.  Each block shows a single training corpus (left) and the evaluation datasets (right) on which that model is tested; numbers are EER (\%).}
  \label{tab:train_eval_matrix}
  \begin{tabular}{llcc}
    \toprule
    \textbf{Train set} & \textbf{Evaluation set} & \textbf{MoLEx} & \textbf{benchmark} \\
    \midrule
    \multirow{5}{*}{ASVSpoof 2019 \cite{wang2020asvspoof}} 
      & ASVSpoof 2019LA & 0.44 & 0.65\textsuperscript{\cite{pan24c_interspeech}} \\[2pt]
      & ASVSpoof 2021LA \cite{liu2023asvspoof} & 4.31 & 3.50\textsuperscript{\cite{pan24c_interspeech}} \\
      & ASVSpoof 2021DF \cite{liu2023asvspoof} & 3.32 & 3.10\textsuperscript{\cite{pan24c_interspeech}} \\
      & In-the-Wild \cite{muller22_interspeech}      & 9.60 & 11.83\textsuperscript{\cite{schafer2024robust}} \\
      & LibriSeVoc \cite{sun2023ai}       & 1.45 & 1.54\textsuperscript{\cite{sun2023ai}} \\
    \midrule
    DFADD \cite{du2024dfadd} & DFADD & 0.0 & 4.01\textsuperscript{\cite{du2024dfadd}} \\
    \midrule
    Fake-or-Real \cite{FoR_dataset} & Fake-or-Real & 0.34 & 1.06\textsuperscript{\cite{speecharena-df-leaderboard}} \\
    \bottomrule
  \end{tabular}
\end{table}

\begin{table*}[!t]
  \renewcommand{\arraystretch}{0.92} 
  \centering
  \caption{Summary of the bona-fide sources and spoof attacks from the experimental datasets. Spoof pipeline lists the dominant attack families.}
  \label{tab:datasets}
  \small
  \begin{tabular}{llll}
    \toprule
    \textbf{Dataset} & \textbf{Bona-fide source} & \textbf{Spoof pipeline} \\
    \midrule
    ASVSpoof 2019 LA   & VCTK \cite{Veaux2017CSTRVC} (107 speakers, studio) & 13 neural TTS/VC (WaveNet, Tacotron2, \emph{etc.}) \cite{van2016wavenet,shen2018natural} & \\
    ASVSpoof 2021 LA   & VCTK (same)             & Same 13 TTS/VC + telephony codec variants         \\
    ASVSpoof 2021 DF   & VCTK (same)             & Same 13 TTS/VC + social-media compression         \\
    ASVSpoof 5 (2024)  & MLS crowdsourced speech \cite{pratap20_interspeech} & $>$30 new neural TTS/VC + adversarial filters \cite{popov2021diffusion,lux2022low,kim2021conditional,kim2020glow,steiner2017creating}     \\
    DFADD              & VCTK                    & Diffusion / flow-matching TTS \cite{kim2023p,mehta2024matcha,li2023styletts,popov2021grad}          \\
    LibriSeVoc         & LibriTTS \cite{zen2019libritts}      &6 GAN-based neural vocoders \cite{yamamoto2020parallel,kong2020diffwave,chen2020wavegrad}            \\
    In-the-Wild        & Celebrity podcasts      & 219 publicly posted deepfakes (unknown TTS)       \\
    Fake-or-Real & Arctic, LJS, VoxForge \cite{voxfoge}   & Deep Voice 3, WaveNet TTS, proprietary TTS  \cite{ping2017deep,shen2018natural}       \\
    \bottomrule
  \end{tabular}
\end{table*}

To verify that the gains obtained on ASVSpoof 5 transfer to corpora with different speakers, recording conditions and spoofing pipelines, we repurpose the same MoLEx configuration ($L=12$, $\mathcal{M}=12$, $K/N = 4/12$) and train it separately on three widely–used audio deepfake datasets: ASVSpoof 2019 LA \cite{wang2020asvspoof}, DFADD \cite{du2024dfadd}, Fake-or-Real \cite{FoR_dataset}.  
Table \ref{tab:train_eval_matrix} summarizes our EERs for various evaluation datasets, with the single-system SOTA numbers reported in the corresponding literature, where multiple variants are available we cite the strongest non–ensemble result.

Table\ref{tab:train_eval_matrix} shows that a single MoLEx configuration generalizes well beyond its source corpus. When trained on ASVSpoof 2019, MoLEx matches the previous SOTA on its native evaluation set ($0.44\%$ vs. $0.65\%$ EER) and, without any finetuning, still outperforms the published single–system baseline on the noisy In-the-Wild audio deepfake collection and LibriSeVoc (9.63\% vs.\ 11.83\%, and 1.45\% vs. 1.54\%).  
Performance slightly degrades on the codec-rich ASVSpoof 2021 tracks and on LibriSeVoc, yet remains competitive. When MoLEx is instead trained on DFADD or Fake-or-Real it achieves improving the benchmark EER from $4.01\%$ to $0.00\%$ on DFADD and from $1.06\%$ to $0.34\%$ on Fake-or-Real.

In particular, we are interested in which LoRA experts the router calls upon when it faces spoofing artifacts that differ in generator type, channel, or speaker pool. We therefore move from cross-dataset scores to a qualitative inspection of the expert utilization heatmaps illustrated in Fig.\ref{fig:all_8_heatmaps}. It reveals a clear pattern: subfigures (a)(b)(c), which correspond to ASVSpoof 2019 and the two ASVSpoof 2021 tracks, show nearly identical expert activation stripes across layers, whereas the remaining datasets produce distinct or more diffuse patterns.  We explain this observation from the sources of these datasets, which are briefly summarized in Tabel \ref{tab:datasets}. The ASVSpoof 2019 and 2021 datasets share the same 107 VCTK speakers and the very same set of 13 WaveNet/Tacotron-style TTS/VC attacks; 2021 merely adds telephony or social-media codecs after synthesis. MoLEx therefore re-uses the same similar “WaveNet/Tacotron specialists’’ for all three benchmarks. In contrast, diffusion-based spoofs in DFADD, vocoder artefacts in LibriSeVoc, and the heterogeneous, noisy material in In-the-Wild, ASVSpoof 5 and Fake-or-Real lie outside that lineage, so the router shifts probability mass to alternative experts or spreads it across all four.  
These utilisation maps thus confirm the intended behaviour: MoLEx conserves parameters by re-activating experts when the spoof mechanisms or speaker pools are similar, and automatically allocates new capacity when they are not.

\subsection{Adapt MoLEx on other dataset domain}

To evaluate the adaptability of MoLEx, we extend a model pre-trained on ASVSpoof 5 to new datasets (ASVSpoof 2019LA, 21LA\&DF), assessing its ability to generalize to unseen attacks while mitigating catastrophic forgetting. A baseline model trained solely on ASVSpoof 5 (config 1) performs poorly on new attacks (34.49\%, 40.34\%, 14.37\% on 19LA, 21LA\&DF), highlighting the need for adaptation. Adding two new LoRA experts per MoLEx (config 2) significantly improves performance on new datasets but leads to some degradation on ASV5. To counteract forgetting, one common approach is to retain a portion of the source domain data \cite{robins1995catastrophic}. Config 3 incorporates 10\% of ASVSpoof 5 data during training, improving overall stability (9.12\%, 0.49\%, 3.40\%, 1.77\% on ASV5, 19LA and 21LA\&DF). Config 4, which expands the number of LoRA experts further, yields only marginal gains. The results demonstrate that MoLEx effectively adapts to new domains by integrating additional LoRA experts while retaining prior knowledge, and a small portion of previous data helps maintain past performance without compromising generalization.

\begin{table}[!t]
\renewcommand{\arraystretch}{0.92} 
\centering
\caption{Performance on MoLEx adaptation experiments. Config 1: baseline model ($L=12$, $\mathcal{M}=12$, $K/N = 4/12$) trained only on ASVSpoof 5.
config 2: freeze the config 1 model, adding 2 new LoRA experts per MoLEx module, trained on ASVSpoof 2019 LA.
config 3: same as Config 2, but 10\% of ASVSpoof 5 train data is included to reduce forgetting.
config 4: Expands config 3 by adding 4 LoRA experts for increased capacity.}
\scriptsize
\footnotesize
\setlength{\tabcolsep}{0.1mm}
\renewcommand{\arraystretch}{1.0} 
\begin{tabular}{c|cccc}
\hline
\toprule
\textbf{Config}\quad & \quad \textbf{ASV5} \quad & \quad \textbf{ASV19} \quad & \quad \textbf{ASV21LA} \quad & \quad \textbf{ASV21DF} \quad\\
\hline
1 & 5.56 & 34.49  & 40.34 & 14.37\\
2 & 13.08 & 0.74  & 4.44 & 4.60 \\
3 & 9.12 & 0.49  & 3.40 & 1.77 \\
4 & 9.63 & 0.57  & 3.18 & 2.00 \\

\hline
\bottomrule
\end{tabular}
\label{tab: MoLEx adaptation}
\end{table}

\section{Conclusion}


We propose MoLEx, a parameter-efficient framework that integrates LoRA with MoE router, thus enabling scalable adaptation while keeping the backbone speech SSL model frozen. On the challenging ASVSpoof 5 benchmark MoLEx consistently surpasses strong single-model baselines, and the proposed orthogonality regulariser further boosts performance by forcing each low-rank expert to exploit its full capacity. Crucially, the expert-utilization analysis shows that MoLEx is domain-aware: the router re-activates the same specialists for attacks drawn from the shared VCTK + WaveNet/Tacotron lineage (ASVSpoof 19/21), yet switches to or blends in other experts when confronted with diffusion-based, vocoder-only or in-the-wild spoofs. Finally, we demonstrate the scalability of MoLEx by re-training only new experts and adapting to new data domains.



\bibliographystyle{IEEEtran}
\bibliography{IEEEabrv,mybib}
\end{document}